\documentstyle{article}
\begin{document}
\begin{center}
\LARGE{Distinquishing of indistinquishable. Coarse graining of relationships.}
\small\footnote{Remarks to the missing science of mind.
Most of the present topic were published in quantum-mind
digest \cite{qmind}}
\end{center}
\begin{center}
\small{Oleg Gorsky}\\
Transmag Research Institute, Dnepropetrovsk, Pisarjevsky 5, Ukraine \\
e-mail:gorsky@npkista.dp.ua
\end{center}

\begin{abstract}
At present many of endeavours in physics are made to recognize  in quantum
substance classical reality, time, subjectivity, consciousness and many other
physical and nonphysical features. The purpose of these remarks is to draw 
attention to a fact that perplex situations are to be 
understood in 'unteared off' manner of fundamental and relative 'mixture'.   
\end{abstract}

The  crucial point in mind-matter division is  coarse-graining
of relationships.
From the beginning I do intend to distinquish coarse-graining
of consistent histories as approach to quantum theory, coarse graining
of observables in thermodynamics and fractal approach
and coarse graining of relationships.
In quantum theory  one does suppose some notion of a coarse
grained configuration space, whose observables correspond to suitable
averages of observables of the physical configuration space [2].
As compare to coarse graining of consistent histories I do name
mind-matter coarse graining as coarse graining of relationships.
To understand what I am speaking about I do recall 'the experiment'
well-known to most of the people.

Let I do remind two trains
moving along one another.In some specific regimes of motion
(rectilinear uniform motion) mind does
lost  stability in distinqushing of its state.This means that
in the same train some observers will  'stand',
some will 'move' and some will
'jump' from stand to move. A relationships of the observers to the situation
will be ambiguous and there are no ways to change the situation.
The situation may be described as relativism ( and fundamental relativism)
of Galileo as next:{\it None of the physical devices does distinquish stasis
from rectilinear uniform motion}. All this is well known. But peculiarity
of the situation is at hand. {\bf Mind does declare indistinquishability
of the states only when mind does distinquish the both ( stasis and
rectilinear uniform motion)!} That is what I do name as coarse graining
of relationships.This is my only task to show that as far as I know
coarse graining of relationships is inherent to mind in most of its work.
I do give some more examples on the subject.

Suppose  I am considering two abstract objects which are equal. Something
I may say about the situation, but what? To be understood
I denote one of an object as A and the second one as B the objects become
equal but distinquishable (I write A=B).If I write A=A or B=B
than I have no possibility to continue 'something to say'.
So A and B are  'and' physically distinctive ( 'A' differs from 'B')
and may be only because I may continue 'something to say'
(for example, if A=B then B=A).
So I may say that 'A' and 'B' are 'markers' too,  along many other
relationships to the situation. This situation may
be treated too as 'distinquishing of indistinquishable' as in two trains
experiment.The same discourse one might succeed if distinquishability
is marked by words.There is very subtle moment,I think, is not to be ignored.
So denotions ( or words) give rise to 'and' physical differentiations
of abstract objects and may be because abstract discourses are possible.
If someone will disagree then I ask 'to tear off' previous remarks and
show the 'place' where it will be always done.I would like to underline
that numbers ( say, one, two, three...) one does distinquish in the words too.
That means that one does distinquish say three from two 'and' 
as informational (symbols) 'and' as physical objects.

Let us take the situation with tossing of classical coin.
In a physical coin two sides are unavoidable physical different ( that is
to distinguish two sides).If one does take a coin with, say, two heads on both
sides then situation does become  perplex. Indeed, in the case an observer in
her/his mind does distinquish both sides but experiment does  show
the indistinquishing. Possibilities are lost,one has strong deterministic
situation but mind 'whispers' that it is false. Of course different
observers will speak different about the situation but nothing does help.
Situation will be perplex and there is no ways to 'untie' the situation.
So if one is considering abstract coin one needs to
justify how this physical difference of the sides of the coin will
be neglected and this justification must be very subtle ( it is
clear that 'and' physical difference may play a role in futher explanation
of behavior as far as dynamical behavior of the coin).
And then: How will one  distinguish the sides of abstract coin? If there are
2 sides then how one may count them if they are not  marked this means
how one may know the number of sides if they are similar?
How does one  know that in abstract coin there are heads and tails
( the essential is not the names or 'colours' or 'markers' but difference
of names or 'colours' or 'markers')? 'Markers'( symbols,words)
begin their life because they are not 'worse' then real heads and tails.
How does one distinquish real heads and tails? 
For present it is suffice to say that both sides in the coin
are distinquishable physically and informatively. So are any 'markers'.
The situation for mind to some degree doesn't change because it is very
hard to justify that real heads and tails are 'better' then real symbols
or words such as 'heads' and 'tails'. Distinquishing of indistinquishable one
may see in many fields.

Now I will try to show how coarse graining of relationships does
play the 'game' within physics. Of course, there are a lot of to show
but at present I'll take the example frequently discussed in thermodynamics
literature.
We often see coffee cup breaking accidently into pieces but never a cup
reform from the pieces.
The question is "Why?" ( let us denote this as $Q_1$). The most conventional
answer is "Because the probability of the event in practice is very small".
But let us get a view as next. Suppose one have seen a cup never before but 
now do
meet a lot of pieces. The question $Q_2$ may be now as "Is it that $Q_1$ 
is possible?"
Yet once more suppose that one does meet a lot of white pieces. 
Is the question $Q_3$
" Why we never have seen papalupu reform from the pieces?" possible?
\small\footnote{Any question is possible simply because I am not restricted
in asking question in the present remarks. It is very hard if possible
to show how asking question in advance is different from
asking question in situations described above.}
I don't know what is papalupu. But the question is not so absurd as it
seems. Why does one not distinquish papalupu in pieces contrary to
the cup?
I doubt that $Q_1,Q_2,Q_3$ may be compatible and answers $A_1,A_2,A_3$
will be consistent within any space-time localization of distinquishing. 
The situation is perplex but might be coarse
grained. May be this is the straight  way to tear  $Q_1,Q_2,Q_3$ off by saying of
standard thermodynamics answer ( as above) and ignoring all other.

Coarse graining of relationships does reveal the relative 'gap'
where causality does fail.There is no physical cause of being 'standing',
'moving' or 'jumping' in two trains experiment.There is no cause of saying
that abstract objects do reveal themselves through things,symbols,
words and else.

Indeed, {\it acausality  does reveal itself as possibilities in perplex
'unteared off' situation through coarse graining}. I am not aware of 
any situation where coarse graining of relationships would not appear 
into strange form in future by saying " We didn't see this before 
but now it is obvious".

In the light of previous remarks the task of creation of artificial mind may be
formulated as follow:{\bf How has one to perform physical device which does
distinquish stasis and rectilinear uniform motion?} This sounds
a bit like paradox.

\end{document}